\documentclass[prb,     twocolumn, letterpaper]{revtex4-2}
\usepackage{afterpage}
\usepackage{xcolor}
\definecolor{myblue}{RGB}{0, 0, 255} 
\usepackage{amsmath}
\usepackage{xr}
\usepackage[colorlinks=true, linkcolor=blue, citecolor=blue, urlcolor=blue]{hyperref}
\usepackage{subfig} 
\usepackage{graphicx} 
\usepackage[format=plain,justification=justified]{caption} 
\usepackage{orcidlink}
\begin{document}
\title{From Structural Stability to Electronic Flexibility: Unveiling Strain-induced Effects in a MoS$_2$/Perylene Orange Hybrid System}
\author{Mohammed El Amine Miloudi \orcidlink{0000-0002-2654-8758}}
\author{Oliver Kühn\textsuperscript{*}\orcidlink{0000-0002-5132-2961}}
\email{oliver.kuehn@uni-rostock.de}
\affiliation{University of Rostock, Institute of Physics, Albert-Einstein-Str. 23-24, D-18059 Rostock, Germany }

\date{\today}
\begin{abstract}

This study delves into the interaction between a monolayer of molybdenum disulfide (MoS$_2$) and a single Perylene Orange (PO) molecule, representative of inorganic and organic semiconductor materials, respectively. Investigation of the amalgamation of these materials under mechanical strains reveals significant alterations in the electronic properties of the MoS$_2$/PO interface. Tensile strain induces a reduction in the bandgap, while compressive strain initially engenders an increase, followed by a subsequent decrease. Notably, MoS$_2$ undergoes a transition from a direct to an indirect bandgap under both stretching and compression conditions. These alterations stem from shifts in the density of states and band structure adjustments resulting from lattice deformations induced by applied strain. Remarkably, under specific compression conditions, the MoS$_2$/PO system manifests a transition between type II and type I band alignments. The detailed analysis of a range of strain magnitudes yields profound insights into the behavior of MoS$_2$ and MoS$_2$/PO systems under mechanical strain, with potential implications for nano- and optoelectronics applications.

\end{abstract}

\maketitle

\section{INTRODUCTION}

The field of advanced optomechanical materials has witnessed substantial growth, particularly with the advent of two-dimensional transition metal dichalcogenides (2D TMDs) \cite{dai19_1805417,peng20_190,ref1}. Among these materials, molybdenum disulfide (MoS$_2$) stands out due to its exceptional optical and mechanical properties \cite{ref2}. Recognized for these properties, MoS$_2$ exemplifies  distinctive features of 2D TMDs, placing them at the forefront of innovation in state-of-the-art optomechanical devices \cite{ref1, ref2, ref3, ref4, ref5}.

An important characteristic of TMDs is their layered structure, having high mechanical stability such as to allow monolayers to sustain up to 10\% strain \cite{ref6}. Applying mechanical strain significantly alters the band structure and optical characteristics of these materials \cite{ref1, ref7}. This  renders TMDs susceptible for applications involving mechanical deformation \cite{ref2}.  In fact, mechanical deformation has been shown to govern the energies of excitons \cite{ref1, ref3, ref8} and phonons \cite{ref9, ref10} in ultrathin semiconductor layers. It leads to alterations in phonon-mediated exciton relaxation rates, changes in the phonon energy spectrum, and variations in the strength of exciton-phonon coupling \cite{ref6, ref11}.

Experimental efforts have largely focused on the highly symmetrical MX$_2$ family, where 'M' denotes metals like Mo or W, and 'X' includes elements such as Se or S \cite{ref1, ref6, ref7}. Previous studies have revealed significant effects upon application of uniaxial deformation to MoS$_2$ \cite{ref2, ref10, ref12}, encompassing changes in electrical mobility, shifts in Raman lines, alterations in photoluminescence (PL) spectra, electron transport characteristics, variations in Schottky barrier heights, bandgap tunings, and the emergence of piezoresistive phenomena \cite{ref10, ref12}.

In addition to mechanical strain effects, recent research has delved into the interaction between transition metal dichalcogenides (TMDs) and organic molecules, leading to the emergence of hybrid TMD/organic interfaces with novel optoelectronic properties \cite{ref13, ref14}. These interfaces represent a synergistic amalgamation of the strengths of both components, offering a diverse array of benefits across various technological applications \cite{ref14, ref15, ref16, padgaonkar20_763a}. While organic molecules are recognized for their outstanding light-absorption capabilities, they may confront challenges related to charge mobility and stability. In contrast, inorganic materials excel in charge transport but may exhibit limited light-absorption capabilities \cite{ref14, ref15, ref16, ref17, ref18}. The introduction of organic molecules onto TMD surfaces serves to enhance crucial properties such as charge carrier mobility and PL quantum efficiency by proficiently passivating defects within the TMD structure and mitigating disruptive electron screening effects \cite{ref16, ref18}.
Furthermore, the integration of organic dye molecules with MoS$_2$ holds promise in augmenting the optoelectronic characteristics of photodetectors. Extensive investigations have explored the synergy between MoS$_2$ and diverse organic molecules, including perylene derivatives, among others, highlighting the pivotal role of the hybrid interface for bolstering charge transfer, and modifying  photoemission \cite{ref14,ref18, ref19, ref20, ref21, ref22, ref23}. This enhancement stems from the intricate interaction between MoS$_2$ and organic molecules, leading to improvements not only in structural integrity but also in the overall optoelectronic performance of the resultant hybrid system.
To further elucidate these advancements, the types of interfaces formed between TMDs and organic molecules can be distinguished as being of  type I or type II band alignment. Such categorization already provides valuable guidance into the underlying mechanisms governing charge transfer processes and exciton dynamics or PL within these hybrid systems.

Perylene Orange (PO), i.e. N,N'-bis(2,6-diisopropyl phenyl)-3,4,9,10-perylene tetracarboxylic diimide (C$_{48}$H$_{42}$N$_2$O$_4$), is a widely studied organic dye molecule \cite{ref24, ref25, ref26, ref27}. Its  molecular structure (cf. Fig. \ref{fig:unique_label2} below) comprises a perylene tetracarboxylic diimide (PTCDI) core, featuring fused benzene rings with four carbonyl and two imide groups. N, N'-bis(2,6-diisopropyl phenyl) substituents are connected to the imide nitrogen atoms, each incorporating an identical 2,6-diisopropyl phenyl group \cite{ref25, ref28, ref29}. The planar and polycyclic aromatic architecture results in a vibrant coloration, establishing PO as a highly sought-after pigment and dye with versatile applications \cite{ref30}. It exhibits exceptional optical characteristics, including robust absorption and emission properties within the visible spectrum. Consequently, it emerges as an optimal choice for applications that require enduring and vivid coloration \cite{ref32, ref33, ref34}.

PO utility extends beyond pigmentation to optoelectronics \cite{ref32, ref33}. Its exceptional properties, including tunable absorption and emission, have garnered interest for developing organic electronic devices \cite{ref29, ref30, ref33, ref34}. The unique combination of high PL quantum yield and efficient charge transport capabilities in PO holds promise for applications such as organic light-emitting diodes (OLEDs). Perylene diimide (PDI) derivatives, similar to  PO, exhibit high PL quantum yields and have proven effective in OLEDs. N-annulated PDI with bulky side chains has demonstrated PL quantum yields over 20\% in the solid state and around 80\% in solution. These materials have been successfully used in OLEDs, achieving significant brightness and low turn-on voltages, making them promising for red-colored OLEDs \cite{dayneko21_933, sabatini19_2954}. Moreover, the rigid structures of these derivatives prevent aggregation, enabling high PL quantum yields at large dye loadings, suitable for room temperature polariton lasing and other advanced optoelectronic applications \cite{sabatini19_2954}.

In this study, we focused on the  MoS$_2$/PO system, elucidating the adsorption geometry, mechanical stability,  and alignment between MoS$_2$ band structure  and PO molecular orbitals. In particular, we investigated the dependence of band alignment on applied mechanical strain. Our findings confirmed a type II band alignment in unstrained MoS$_2$/PO. Application of strain is shown to  allow for a  precise control of the bandgap and alignment, transitioning to type I under specific conditions. This control provides direct influence over charge transfer dynamics and photoemission characteristics. Our work shows the potential for fine-tuning these attributes in the MoS$_2$/PO system, opening up potential flexibility in designing hybrid systems for the possibility of breakthroughs in nano- and optoelectronics.

\begin{figure}[h]
  \centering
  \includegraphics[width=0.5\textwidth]{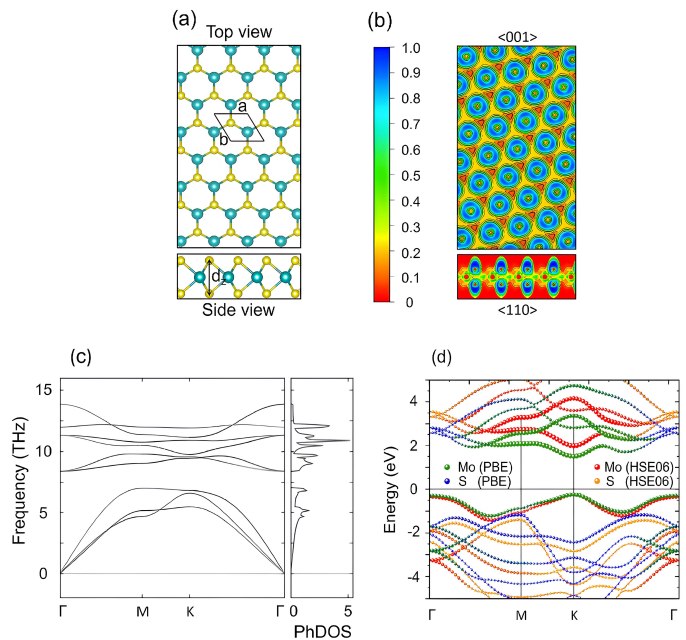}
  \caption{(a) Top and side  views of 2H-MoS$_2$, where large and small spheres represent Mo and S atoms, respectively. The figure illustrates a supercell with the unit cell denoted by a black box. The symbols "a" and "b" denote the lattice constants. (b) The isosurface of the electron localization in the $\langle 001 \rangle$ and $\langle 110 \rangle$ planes, with a contour interval of 0.1. (c) Phonon dispersion and density of states (PhDOS), and (d) band structures obtained using the PBE and HSE06 methods for the MoS$_2$ unit cell (energies given with respect to the Fermi energy $E_{\rm F}$). }
  \label{fig:unique_label}
\end{figure}

\section{COMPUTATIONAL DETAILS}

First-principles calculations were conducted using the Vienna \textit{ab initio} simulation package (VASP), employing projector augmented wave (PAW) pseudopotentials and the Perdew–Burke–Ernzerhof (PBE) exchange-correlation functional \cite{ref36, ref37, ref38}. To prevent interactions between periodic layers, a 20 Å vacuum spacing was included. Additionally, to address the influence of dipole moments, a $z$-axis dipole correction was used. The kinetic energy cutoff was set at 450 eV after confirming convergence, and $\Gamma$-centered grids with dimensions of $17\times17\times1$ and $1\times1\times1$ were used for atomic relaxation of the MoS$_2$ unit cell and the  MoS$_2$/PO supercell, respectively. A denser grid was adopted for the density of states (DOS) and band structure calculations.
The electronic convergence criterion was set at $10^{-6}$ eV, while for geometry optimization the maximum force components were converged to levels below $10^{-2}$ eV/Å.
 To account for long-range van der Waals (vdW) interactions, the DFT+D3 method was incorporated into atomic relaxation and total energy computations, effectively including pairwise electron density interactions with empirically determined parameters \cite{ref39}. The valence electron configurations considered in these calculations are $4d^5 5s^1$ for Mo, $3s^2 3p^4$ for S, $2s^2 2p^2$ for C, $2s^2 2p^3$ for N, $2s^2 2p^4$ for O, and $1s^1$ for H.

Electronic band structures of the MoS$_2$ unit cell were performed using both the PBE \cite{ref40} and the Heyd-Scuseria-Ernzerhof (HSE06) hybrid functional methods, with a standard mixing parameter of 0.25 employed exclusively for the HSE06 calculations \cite{ref41}. This comparative calculation was performed to verify the consistency of the bandgap nature and its position. For the MoS$_2$/PO supercell only PBE calculations were performed. The MoS$_2$ supercell consists of 275 atoms, while the MoS$_2$/PO system comprises 339 atoms. Note that we consider the case of low PO coverage only, i.e. a single PO molecule per supercell.  The phonon dispersion was determined utilizing a $2 \times 2 \times 1$ supercell and density functional perturbation theory (DFPT) \cite{ref42}.

Molecular orbital and frequency calculations for the isolated PO were performed using the PBE functional with a 6-31G(d) basis set as as implemented in the QChem code \cite{epifanovsky21_084801}.

\section{RESULTS AND DISCUSSION}

\subsection{Isolated MoS$_2$ and PO}

Before presenting an in-depth exploration of the hybrid MoS$_2$/PO system, some results on the pristine 2H-phase MoS$_2$ monolayer are summarized in  {Fig.~\ref{fig:unique_label}}. Panel (a) shows the unit as well as  the supercell. Panel (b) provides the electron localization function (ELF), which is employed to qualitatively analyze the nature of chemical bonding within the material. The Mo cation sites exhibit low ELF values, while the S sites display prominent blue annular regions. This significant contrast in ELF values indicates an ionic nature of the chemical bonding. However, the bonding also exhibits covalent characteristics, as evidenced by the non-spherically symmetric high ELF regions around the S anions, which form lobes directed towards the Mo ions. Additionally, there are lobes observed between adjacent S anions, both within the same layer and between two layers. Therefore, the bonding in MoS$_2$ is a hybrid, consisting of both ionic and covalent components, with a slight covalent interaction present between adjacent S anions. 

In  {Fig.~\ref{fig:unique_label}(c)} the phonon dispersion along the high symmetry path ($\Gamma-X-M-\Gamma$) within the Brillouin zone (BZ) is given for the unit cell. We also show the corresponding phonon density of states (PhDOS). Given these results, we note that throughout the BZ, no instances of imaginary phonon frequencies were observed, thereby confirming the dynamic stability of the model used for the  primitive MoS$_2$  \cite{ref43}. Calculations of bond lengths reveal Mo-S, Mo-Mo, and $d_z$ (S-S) distances measuring 2.41, 3.16, and 3.13~\AA, respectively. These values are in good  agreement with previous experimental and theoretical findings \cite{ref44, ref45, ref46, ref47, ref48}. 

The electronic band structure of the unit cell is plotted in   {Fig.~\ref{fig:unique_label}(d)}, confirming the well known direct bandgap at the $K$ point \cite{lebegue09_115409,mak10_136805,splendiani10_1271}.
As far as the functional dependence is concerned, bandgap values of 1.65 eV and 2.1 eV are obtained for PBE and HSE06, respectively. These findings align excellently with prior theoretical investigations and provide compelling evidence of a strong covalent bond formation incorporating the S-3p and Mo-4d states, thus underscoring the presence of robust hybridization within the monolayer structure \cite{ref43, ref44, ref45, ref46, ref49}.

\begin{figure}[h]
  \centering
  \includegraphics[width=0.5\textwidth]{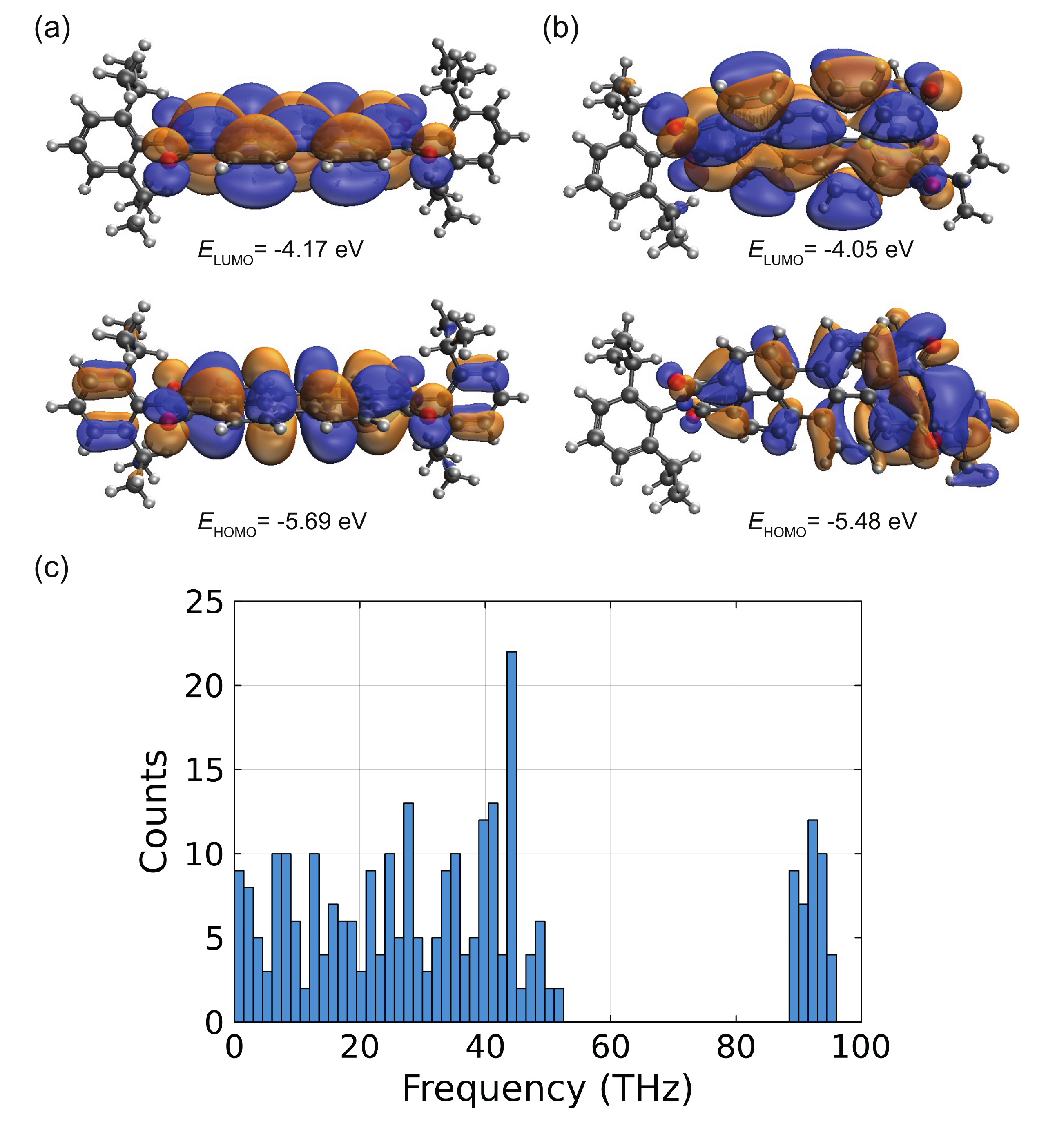}
  \caption{Comparison of PO HOMO and LUMO for  gas phase (energies with respect to vaccum level) (a) and adsorption (b) geometries. Panel (c) shows the distribution of vibrational frequency (histogram) of PO at gas phase geometry.}
  \label{fig:PO-iso}
  \end{figure}

\begin{figure*}[t]

  \includegraphics[width=0.9\textwidth]{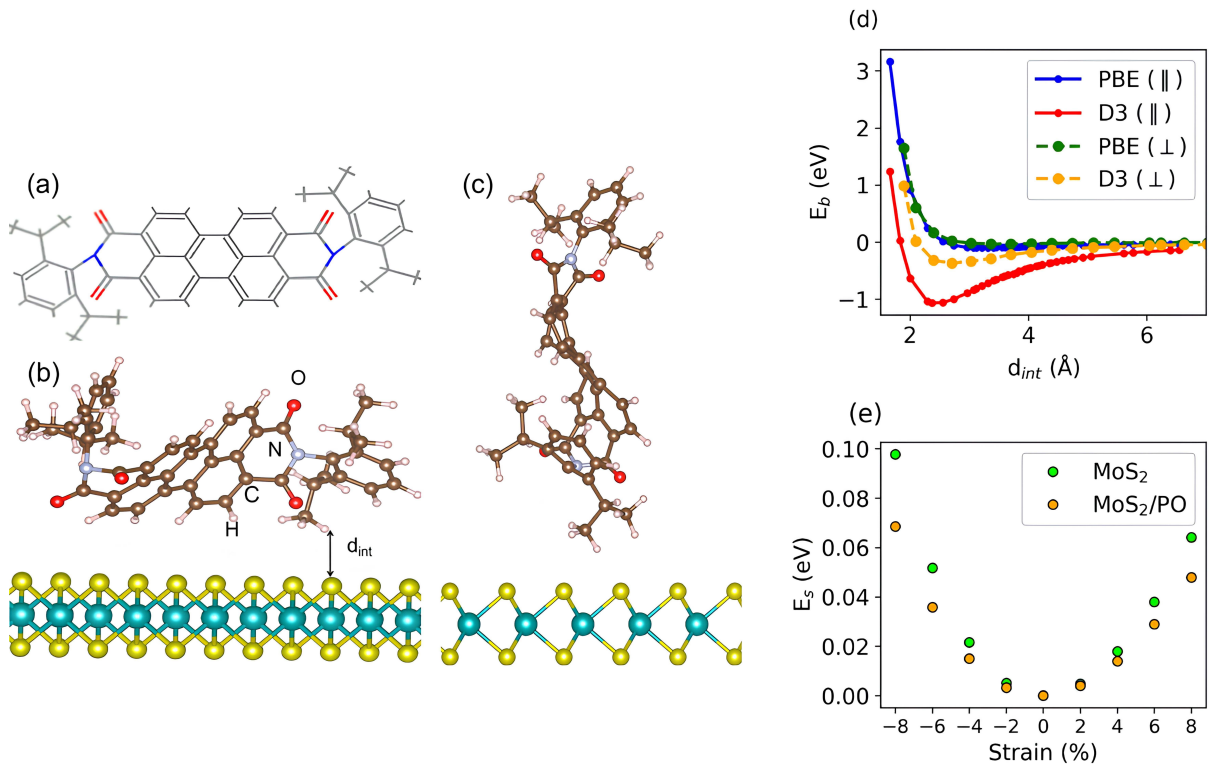}
    \centering
  \caption{(a) Molecular structure of perylene orange (PO), (b) parallel and (c) perpendicular optimized configurations of MoS$_2$/PO hybrid structures, (d) binding energy analysis using PBE and PBE+D3 methods, and (e) comparison of potential energy variation with uniaxial strain in MoS$_2$ and MoS$_2$/PO.}
  \label{fig:unique_label2}
\end{figure*}

Molecular orbitals of PO in gas phase are shown in Fig. \ref{fig:PO-iso}. Specifically, panel (a) and b) show the HOMO (highest occupied molecular orbitals) and LUMO (lowest unoccupied molecular orbital) for the optimized gas phase geometry and the geometry adopted upon adsorption (see also Fig. \ref{fig:unique_label2}). It is noteworthy that for the adsorption geometry the HOMO-LUMO gap changes marginally from 1.52 eV to 1.43 eV only, in particular the HOMO becomes distorted and more localized. In Fig. \ref{fig:PO-iso}(c) the distribution of vibrational mode frequencies of PO is shown for the gas phase geometry. Upon adsorption it is to be expected that in particular the low-frequency part (i.e. below 20 THz) is modified, although not qualitatively. Hence there will be substantial overlap with the bare MoS$_2$ phonon spectrum in Fig. \ref{fig:unique_label}(c).
\subsection{Structural stability and mechanical properties}
A comprehensive exploration was undertaken to characterize the interaction between a monolayer of MoS$_{2}$ and a single PO molecule.  The molecular structure of the molecule is illustrated in  {Fig.~\ref{fig:unique_label2}(a)}. First geometry and binding energy of  both parallel ($ \parallel $) and perpendicular ($\perp $) orientations were scrutinized. Visual representations of the relaxed, i.e. geometry-optimized, atomic structure of the MoS$_{2}$/PO system in these orientations, parallel ($ \parallel $) and perpendicular ($\perp $), can be found in  {Fig.~\ref{fig:unique_label2}(b)} and  {(c)}, respectively. In passing we note that due to the presence of the bulky side groups, the structure of PO distorts substantially at the surface as compared with the isolated molecule case where the perylene core is essentially planar.

The binding energy ($E_{\rm b}$) of the MoS$_2$/PO hybrid interface was calculated for different distances between MoS$_2$ and PO  according to the equation
\begin{equation}
E_{\rm b}=E_{\text{MoS$_2$/PO}}-E_{\text{MoS$_2$}}-E_{\text{PO}}\,.
\end{equation}

Here, $E_{\text{MoS$_2$/PO}}$ is energy of the MoS$_{2}$/PO composite, while $E_{\text{MoS$_2$}}$ and $E_{\text{PO}}$ denote the total (geometry-optimized) energies of the isolated  MoS$_{2}$ and the PO, respectively. Determining the most stable configuration and the equilibrium interlayer distance ($d_{\rm int}$) is based on finding the minimum of $E_{\rm b}$.

In  {Fig.~\ref{fig:unique_label2}(d)} the binding energy is plotted as a function of interlayer distance for the MoS$_{2}$/PO hybrid interface. It is clear from this figure that the parallel configuration using PBE+D3 is more stable than the perpendicular one with $d_{\rm int}$ of 2.37 \AA. In the following, therefore, only the parallel configuration is considered  when exploring the electronic structure and the role of strain. We also note from  {Fig.~\ref{fig:unique_label2}(d)} that in both cases binding is essentially facilitated by dispersion interaction (vdW) and thus of physisorption type.

\begin{figure*}[t]
  \includegraphics[width=0.9\textwidth]{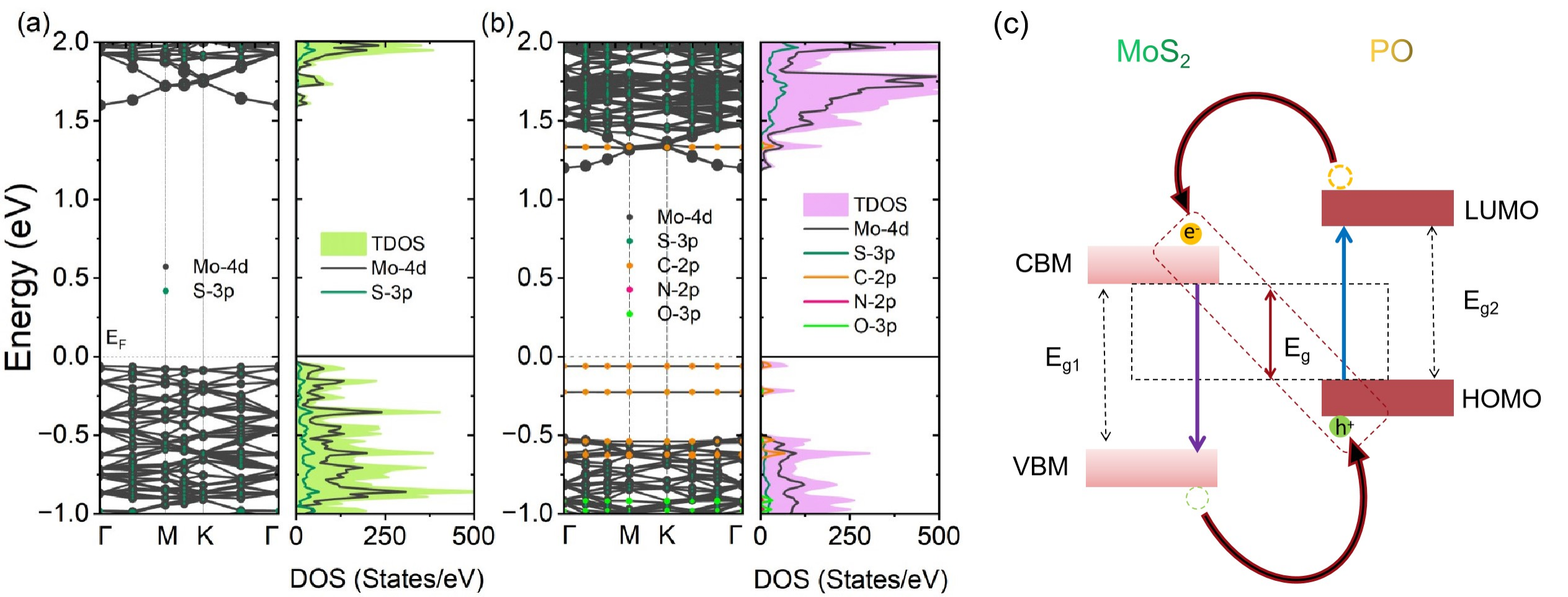}
  \caption{Comparative analysis of band structures and T(P)DOS in (a) MoS$_2$ and (b) MoS$_2$/PO systems, with (c)  schematic of interface band alignment (energies given with respect to Fermi energy $E_{\rm F}$). }
  \label{fig:unique_label3}
\end{figure*}

Beyond structural stability, we studied the mechanical properties of both the MoS$_{2}$ and the MoS$_{2}$/PO systems. Our focus centered on the analysis of the elastic coefficients, specifically $C_{11}$, $C_{12}$, and $C_{22}$. These coefficients were extracted by fitting the unit cell energy, $U$, for a certain set of strain values ($\epsilon_{11}$, $\epsilon_{22}$).

The elastic stiffness coefficients were computed via the following expressions (Note that for the underlying hexagonal lattice it holds that $C_{11}=C_{22}$.)

\begin{equation}
    \begin{aligned}
    C_{11} = \frac{1}{A_{0}}\frac{\partial^2 U}{\partial \epsilon_{11}^2}, \\
    C_{12} = \frac{1}{A_{0}}\frac{\partial^2 U}{\partial \epsilon_{11} \partial \epsilon_{22}},
    \end{aligned}
\end{equation}

with  $A_0$ being the cell area at zero strain. Subsequently, the the Young modulus ($Y$), shear modulus ($G$), Poisson's ratio ($\nu$), and the bulk modulus ($K$) specific to the 2D nature of these materials were derived through the following relationships:

\begin{eqnarray}
Y&=& \frac{C_{11}^2 - C_{12}^2}{C_{11}},\quad  G = \frac{{C_{11} - C_{12}}}{2}, \nonumber\\ \nu &=& \frac{C_{12}}{C_{11}}, \quad K = \frac{{C_{11} + C_{12}}}{2}.
\end{eqnarray}

The analysis involved the systematic application of uniaxial strains $\epsilon_{11}$ and $\epsilon_{22}$ along two in-plane directions, spanning from -0.08 to +0.08, with a step size of 0.02 (for geometrical parameters see Tab. S1 in the Supplementary Material (SM)). To assess the adherence of all considered strains to the elastic limit, the per-atom strain energy, denoted as $E_{\rm S}$, is computed as the difference between the energy of the strained and  unstrained state, normalized by the total number of atoms within the cell. (Notice that the MoS$_{2}$ results have been obtained for the unit cell. A comparative discussion of the differences between unit and supercell is given in the SM, Figs. S1 and S2.) 
The results presented in  {Fig.~\ref{fig:unique_label2}(e)} consistently demonstrate a smooth variation of $E_{\rm S}$ with strain for both MoS$_{2}$ and MoS$_{2}$/PO. From this figure, one also notices that for larger strain values the harmonic limit is no longer valid. Upon compression repulsive interactions give rise to a steeper increase of $E_{\rm S}$, whereas bonds are weakened upon stretching, thus leading to a slower increase of $E_{\rm S}$ as compared with the harmonic approximation. The values for the elastic stiffness coefficients reported below were obtained by fitting in the range $\pm 4$\%. Fits for different ranges are given in the SM, Table  {S2}.

The mechanical properties are summarized in Table~\ref{tab:supp_tableI}. Significantly, both the MoS$_2$ and the MoS$_2$/PO interface satisfactorily meet the Born criteria for a hexagonal structure, wherein $C_{11}$ exhibits a positive value, and $C_{11}$ exceeds $|C_{12}|$. These observations indicate the inherent mechanical stability of both materials.
In particular, the MoS$_2$ demonstrates elastic stiffness coefficients of $C_{11} = 138.49$ N/m and $C_{12} = 31.37$ N/m, which are consistent with previous results \cite{ref50}. For the MoS$_2$/PO hybrid interface, the values are $C_{11} = 140.90$ N/m and $C_{12} = 18.17$ N/m.


\begin{table*}[ht]
\caption{Mesh parameter ($a$), interlayer distance ($d_{\text{int}}$), elastic constants ($C_{11}$, and $C_{12}$), Young modulus ($Y$), shear modulus ($G$),  Poisson’s ratio ($\nu$), and bulk modulus ($K$) of MoS$_2$ and MoS$_2$/PO systems.}
\begin{ruledtabular}
\begin{tabular}{lcccccccccc}
  Material & $a$(\AA) &  ${d_{\rm int}}$  (\AA) & $C_{11}$ (N/m) & $C_{12}$ (N/m)  & $Y$ (N/m) & $G$ (N/m) &  $\nu$  & $K$ (N/m)  \\
  \hline
  MoS$_2$ & 3.18 & -- & 138.49 & 31.37 & 131.38 & 53.56 & 0.23 & 85.24      \\
  MoS$_2$/PO & 3.17 & 2.37 & 140.90 & 18.17 & 138.55 & 61.36 & 0.13  & 79.53  \\
\end{tabular}
\end{ruledtabular}
\label{tab:supp_tableI}
\end{table*}


Table~\ref{tab:supp_tableI} also reflects that both MoS$_2$ and MoS$_2$/PO systems manifest a high Young's modulus and a low Poisson's ratio, symbolic of their remarkable stiffness, incompressibility, and resistance to shear deformation, supported by their shear modulus. The relatively incompressible nature of the hybrid interface can be attributed to the presence of MoS$_2$, having a high bulk modulus. Furthermore, the anisotropic behavior of the hybrid interface is influenced by the organic molecule. An important observation  is the discernible 5\% variance, in Young’s modulus  when comparing MoS$_2$ with its MoS$_2$/PO hybrid counterpart. This deviation underscores the distinct mechanical behavior of the hybrid interface which calls for further investigation, i.e. using Raman or Brillouin spectroscopy to elucidate the underlying molecular and structural dynamics associated with this difference in mechanical rigidity. This difference in Young’s modulus indicates an altered atomic configuration of the MoS$_2$ layer within the hybrid material. Indeed the value of root mean square deviation between isolated MoS$_2$ and  MoS$_2$/PO was  calculated as 0.28 \AA. This small value is in accord with the 5\% change of $Y$ and the notion of physisorption. 

\subsection{Electronic Properties}
%
Results of the  investigation of electronic properties of the MoS$_2$ supercell and its hybrid interface with PO are shown in  {Figs.~\ref{fig:unique_label3}(a)} and  {(b)}, respectively. The left and right panels of these figures give  the band structure and the total/partial (T/P)  DOS. Notice that when performing MoS$_2$ supercell simulations Brillouin zone folding occurs, causing a shift in the direct bandgap from the $K$  to the $\Gamma$ symmetry point (see  {Fig.~\ref{fig:unique_label3}(a)}). In case of pristine MoS$_2$ this is, of course, an artefact of the used supercell. Although this artefact could be removed by band unfolding techniques  \cite{krumland21_044003}, for the sake of comparison with the  MoS$_2$/PO hybrid system we will use the folded band structure for pristine MoS$_2$.

In  {Fig.~\ref{fig:unique_label3}(a)}, the  electronic properties of MoS$_2$ around the bandgap are governed by substantial contributions  from the 4d orbitals of Mo atoms and the 3p orbitals of S atoms. This attribution is substantiated by the respective peaks observed in both the DOS and band structure, confirming in particular  the   hybridization between these orbitals at the Valence Band Minimum (VBM).

Figure~\ref{fig:unique_label3}(b) shows band structure and  DOS for the 
 MoS$_2$/PO hybrid system. 
 Due to the physisorption nature of the adsorption, band gaps of the constituents can still be identified, in addition to the overall bandgap. 
In particular we can assign the  bandgap $E_{\rm g1}$ which characterizes the gap between  VBM and Conduction Band Minimum  (CBM) of MoS$_2$ and  the bandgap $E_{\rm g2}$ delineating the energy difference between the HOMO and LUMO of the PO molecule.
Whereas the CBM is of Mo-4d origin, in the range of the VBM there are contribution of both Mo-4d and C-2p orbitals. States close to the Fermi energy are of molecular origin (mostly C-2p). In fact even the HOMO-1 is well-separated within the MoS$_2$ bandgap.

\begin{figure}[h]
  \centering
  \includegraphics[width=0.4\textwidth]{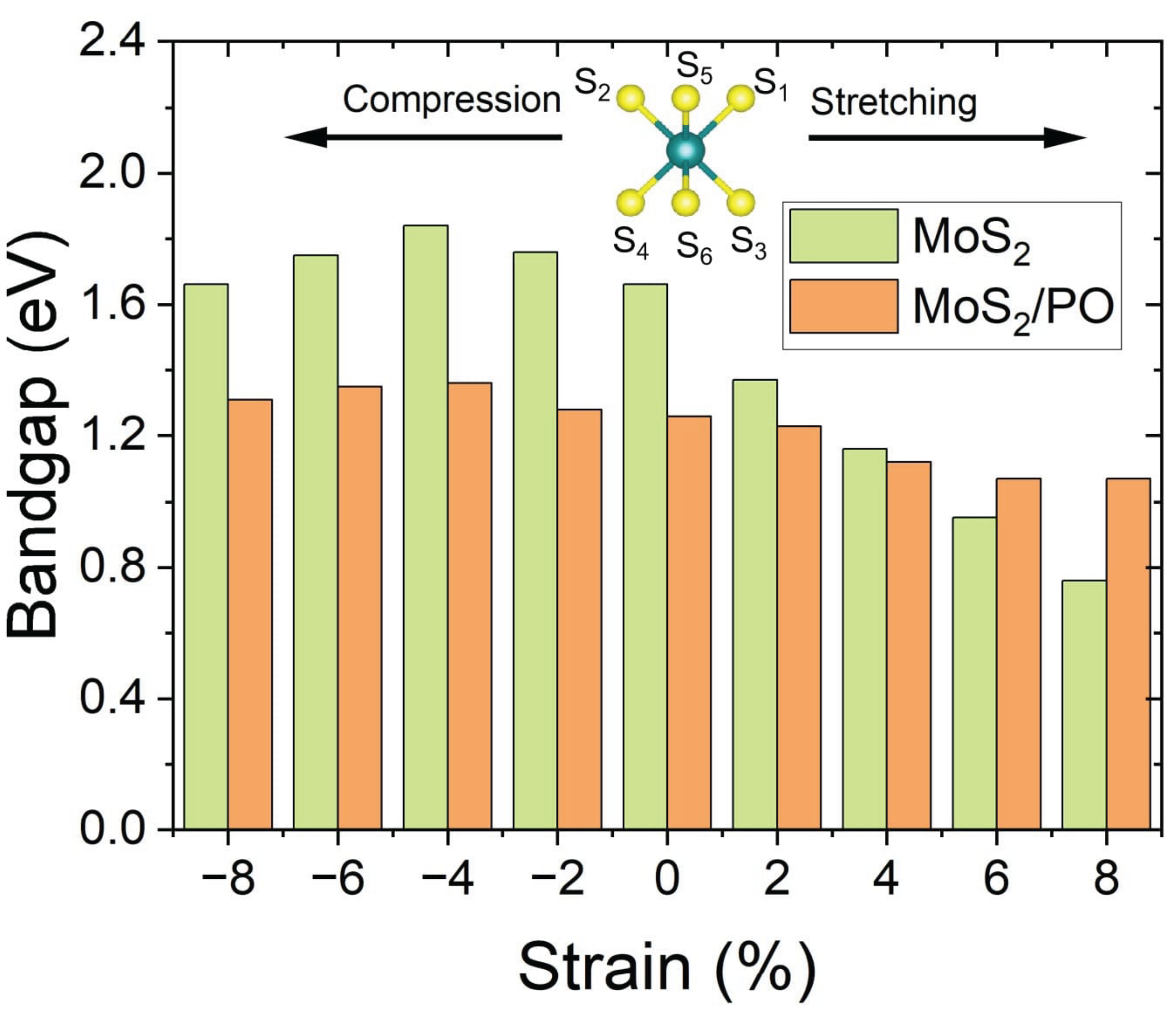}
  \caption{Bandgap ($E_{\rm g}$) evolution in strained MoS$_2$ and MoS$_2$/PO.}
  \label{fig:unique_label5}
  \end{figure}

\begin{figure*}[t]
  \centering
  \includegraphics[width=1\textwidth]{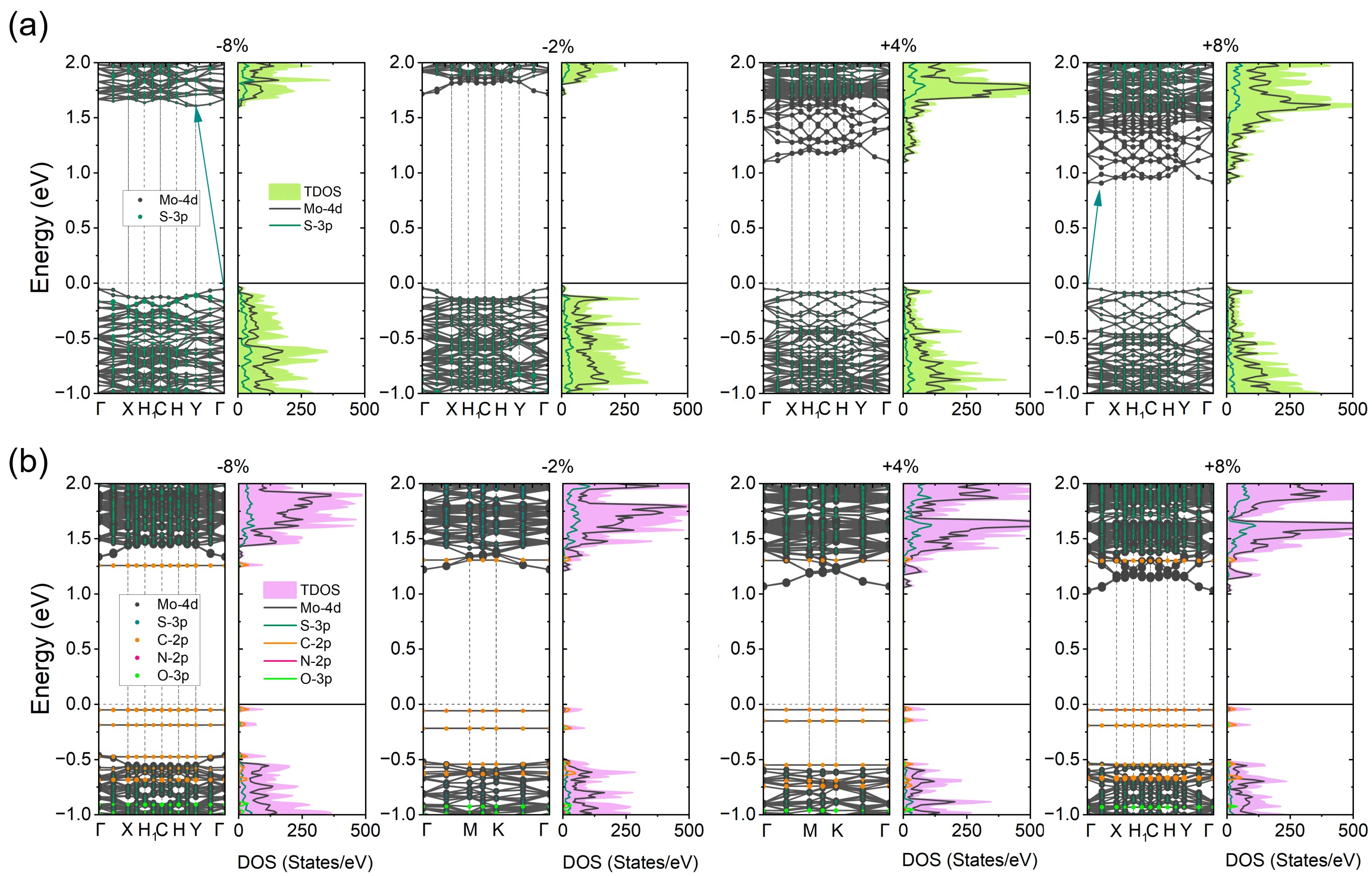}
  \caption{Analysis of band structure and T(P)DOS in  (a) MoS$_2$ and (b) MoS$_2$/PO under varying uniaxial strain (energies given with respect to Fermi energy $E_{\rm F}$). }
  \label{fig:unique_label7}
\end{figure*}

The relative positions of the states forming the two mentioned bandgaps is highly relevant for the optical properties of the hybrid system. The present case can be categorized as type II band alignment, cf. Fig.~\ref{fig:unique_label3}(c). Here upon photoexcitation, electrons in the HOMO and HOMO-1 of PO can be excited to its LUMO. These electrons can then transfer to the CBM of MoS$_2$, leaving behind holes in the PO molecule. This process effectively separates the charge carriers, reducing the likelihood of radiative electron-hole recombination and thus enhancing charge separation efficiency. In principle, the holes in the PO molecule HOMO and HOMO-1 can interact with the VBM of MoS$_2$, potentially leading to additional charge transfer dynamics. Of course, this qualitative discussion should be taken with caution since it completely neglects many body effects.

Such efficient charge transfers would have consequences for the  photoluminescence Mo$S_2$ or the fluorescence of PO, which would be quenched due to the effective separation of electron-hole pairs and the suppression of their recombination. This charge separation is confirmed by  recent experiments on this system, which showed completely quenched PO fluorescence \cite{ref21}.


\begin{table}[!t]
\caption{Strain-dependent electronic characteristics of MoS$_2$: bandgap ($E_{\rm g}$), its type, and position at different strains.}
\begin{ruledtabular}
\begin{tabular}{cccc}
  Strain (\%) & $E_{\rm g}$ (eV) & $E_{\rm g}$ type &  $E_{\rm g}$ position \\
\hline
  -8 & 1.66 & Indirect & [$\Gamma$]-[$Y$] \\
    -6 & 1.75 & Indirect & [$\Gamma$]-[$Y$] \\
    -4 & 1.84 & Indirect & [$\Gamma$]-[$Y$] \\
    -2 & 1.76 & Direct & [$\Gamma$]-[$\Gamma$] \\
 Pristine & 1.65 & Direct & [$\Gamma$]-[$\Gamma$] \\
    +2 & 1.37 & Direct & [$\Gamma$]-[$\Gamma$] \\
    +4 & 1.15 & Direct & [$\Gamma$]-[$\Gamma$] \\
    +6 & 0.95 & Indirect & [$\Gamma$]-[$X$-$\Gamma$] \\
    +8 & 0.76 & Indirect & [$\Gamma$]-[$X$-$\Gamma$] \\
\end{tabular}
\end{ruledtabular}
\label{tab:supp_tableII}
\end{table}

\begin{table}[!t]
\caption{Strain-dependent electronic characteristics of MoS$_2$/PO systems: $E_{\rm g1}$, $E_{\rm g2}$, bandgap ($E_{\rm g}$),  and band alignment type. For an analysis of the effect of geometry alone on the HOMO-LUMO gap, see Table S3 in the SM.}
\begin{ruledtabular}
\begin{tabular}{cccccc}
Strain (\%) & $E_{\rm g1}$ (eV) & $E_{\rm g2}$ (eV) & $E_{\rm g}$ (eV) &     type  \\
\hline
 -8 & 1.78 & 1.31 & 1.31 &  I \\
 -6 & 1.78 & 1.35 & 1.35 &  I \\
 -4 & 1.76 & 1.37 & 1.36 &  II  \\
 -2 & 1.72 & 1.37 & 1.28 &  II \\
 Pristine & 1.71 & 1.38 & 1.25 &   II \\
 +2 & 1.70 & 1.40 & 1.23 &  II  \\
 +4 & 1.66 & 1.35 & 1.12 &   II \\
 +6 & 1.62 & 1.34 & 1.07 &   II \\
 +8 & 1.57 & 1.35 & 1.07 &  II \\
\end{tabular}
\end{ruledtabular}
\label{tab:supp_tableIII}
\end{table}

\subsection{Effect of strain on electronic properties}

Tailoring electronic properties through external manipulation, such as mechanical strain, offers great promise for advancements in the fields of nano- and optoelectronics. In what follows, we explore the   electronic characteristics of the MoS$_2$ supercell and the MoS$_2$/PO hybrid interface under the influence of uniaxial strain. 

 {Figure~\ref{fig:unique_label5}} presents a summary of the predicted bandgap evolution ($E_{\rm g}$) with strain, utilizing the PBE approximation for both MoS$_2$ and MoS$_2$/PO systems. This figure reveals a discernible trend: when subjected to tensile strain, both the MoS$_2$ and MoS$_2$/PO systems   exhibit a systematic reduction of their bandgap. Conversely, as compressive strain increases, reaching a critical threshold of approximately -4\%, the bandgap  decreases with further increasing of the  compressive strain. Our findings regarding MoS$_2$ demonstrate a good agreement with prior experimental results \cite{ref4}.

The  focus of this study centers on  {Fig.~\ref{fig:unique_label7}}, which in panel (a) offers a   detailed analysis of the band structure, as well as the partial and total DOS of MoS$_2$ under varying magnitudes of uniaxial strain. Pristine MoS$_2$ shows a direct $\Gamma \rightarrow \Gamma $ bandgap which is also present for small strain values. At strain values $\le-4$\% and $\ge6$\% the band structure features an indirect bandgap. 
Under compressive deformation conditions, the VBM localizes at the $\Gamma$ symmetry point, while the CBM is positioned at the $Y$ point.  
Conversely, under +6\% and +8\% strains the VBM and CBM positions converge to $\Gamma$ and ($\Gamma$-$X$), respectively. This distinct bandgap behavior, observed under both compressive and tensile strains, significantly deviates from the direct bandgap characteristics exhibited under all other uniaxial strain percentages. Details regarding the bandgap energies value, type, and specific positions under different strain conditions are compiled in Table~\ref{tab:supp_tableII}.

The origin for the observed behavior can  be traced to transformative effects of uniaxial strain on the 2D Bravais lattice of a MoS$_2$ monolayer. In its pristine state, MoS$_2$ has a hexagonal lattice. The application of uniaxial strain, whether tensile or compressive, prompts a deformation of this hexagonal lattice, leading to shifts in the interatomic distances and the angles between lattice vectors (refer to Table \textcolor{myblue}{S4}). As a result, the once-hexagonal lattice assumes an oblique lattice structure

Finally, we point out that,  across all  strain conditions, the dominant contributions in both the valence and conduction bands emanate from the Mo-4d and S-3p orbitals. This unvarying pattern underscores the  significance of these particular orbitals in shaping the electronic landscape of MoS$_2$ across a spectrum of strain magnitudes.

 Fig.~\ref{fig:unique_label7}(b) presents the band structure and DOS for the MoS$_2$/PO system associated with the behavior in Fig. \ref{fig:unique_label5}; specific numbers can be found in Tab. \ref{tab:supp_tableIII}. 
 Comparing Tabs. \ref{tab:supp_tableII} and \ref{tab:supp_tableIII} we notice that the total band gap variation is about 59\% and 21\% of its maximum value for MoS$_2$ and MoS$_2$/PO, respectively. The maximum value is attained at -4\% strain. Inspecting the separate bandgaps, $E_{\rm g1}$ and $E_{\rm g2}$, the variation is 12\% and 6\%, respectively. For the CBM-VBM gap the maximum shifts to large strain values (-6\%), whereas the largest HOMO-LUMO gap is observed at +2\%.
 Notably, the Mo-4d and S-3p orbitals of MoS$_2$ play a substantial role in shaping the  CBM and VBM of MoS$_2$, while the C-2p orbital is pivotal in the formation of the HOMO and LUMO across all examined systems, both strained and unstrained. Further the type of bandgap does not change, i.e. it always features a direct transition.

 Comparing the evolution of the bandgaps with strain one should take into account that due to the physisorptive binding the PO molecule is only weakly affected by strain. Note that the situation would change upon increasing the coverage, i.e. when PO molecules would interact. Thus it is essentially the MoS$_2$ band structure that is shifting relative to the PO MOs, with the HOMO  positioned close to the Fermi energy. Still, the variation of $E_{\rm g1}$ in MoS$_2$/PO is only 20\% of that in MoS$_2$, i.e. the hybrid interface makes the MoS$_2$ band structure less susceptible to strain. 

 Pristine MoS$_2$/PO has a type II band alignment, which doesn't change when applying positive strain. However, the increase of the band gap $E_{\rm g1}$ with increasing negative strain leads to a situation where the LUMO shifts into the MoS$_2$ band gap, i.e. for strain values exceeding -6\% our results predict a change to type I band alignment.
Upon photoexcitation this would enable efficient electron transfer from the CBM to the LUMO, impacting PL in this system.
\section{CONCLUSION}
Using Density Functional Theory we have investigated PO physisorbed onto monolayer MoS$_2$, putting emphasis on the mechanical and electronic properties of this hybrid system. We found that binding of PO is preferentially in parallel (on-top) orientation and facilitated by van der Waals interaction. Upon binding the structure of PO is substantially distorted, without considerably affecting the HOMO-LUMO gap. 

MoS$_2$/PO is mechanically stable with key parameters changing only modestly as compared with bare MoS$_2$. For instance Young's module increases by about 5\%. The mechanical stability suggested to explore the influence of strain on the electronic properties of the hybrid system. As a consequence of the non-covalent binding as well as the low-coverage assumption, molecular orbitals of PO are only marginally affected by strain. But even the band gap evolution of the MoS$_2$ part of the hybrid system with strain is less pronounced in bare MoS$_2$. Still the combination of strain effects on the constituents of the hybrid interface resulted in a qualitative change of the electronic properties, i.e. a switch between type II and type I band alignment upon compressive strain, with all its consequence for the photophysical properties.

The present investigation advances our comprehension of the electronic properties and structural behavior of MoS$_2$ vs. MoS$_2$/PO under mechanical strain, furnishing insights for future endeavors into novel device applications.

\begin{acknowledgments}
This work was funded by the Deutsche Forschungsgemeinschaft (DFG, German Research Foundation) - SFB 1477 "Light-Matter Interactions at Interfaces", project number 441234705”. We thank  T. Korn and J. Schr\"oer (University of Rostock) for their insightful comments on the manuscript.
\end{acknowledgments}

%

\end{document}